\documentclass[twocolumn,showpacs,prl]{revtex4}
\usepackage{dcolumn}
\usepackage{graphicx}
\usepackage{graphicx,amssymb,amsmath,revsymb}
\usepackage{natbib}
\usepackage{multirow}

\newcommand{\diffexp}{a}		
\newcommand{\msdexp}{{b_{\rm M}}}
\newcommand{\tauexp}{{b_{\tau}}}

\begin{document}

\title{Critical scaling and aging in cooling systems near the jamming transition}

\author{David A. Head$^{1,2}$}

\affiliation{$^{1}$Institute of Industrial Science, University of Tokyo, Meguro-ku, Tokyo 153-8505, Japan}
\affiliation{$^{2}$Institut f\"ur Festk\"orperforschung, Theorie II, Forschungszentrum J\"ulich 52425, Germany}

\date{\today}

\begin{abstract}
We conduct athermal simulations of freely--cooling, viscous soft spheres around the jamming transition density $\phi_{J}$, and find evidence for a growing length $\xi(t)$ that governs relaxation to mechanical equilibrium. $\xi(t)$ is manifest in both the velocity correlation function, and the spatial correlations in a scalar measure of local force balance which we define. Data for different densities $\phi$ can be collapsed onto two master curves by scaling $\xi(t)$ and $t$ by powers of $|\phi-\phi_{J}|$, indicative of critical scaling. Furthermore, particle transport for $\phi>\phi_{J}$ exhibits aging and superdiffusion similar to a range of soft matter experiments, suggesting a common origin. Finally, we explain how $\xi(t)$ at late times maps onto known behavior away from $\phi_{J}$.
\end{abstract}



\maketitle

%
%

Discontinuous phases such as granular media, foams and emulsions often exhibit a {\em jamming} transition from a fluid to an amorphous solid as the volume fraction is increased~\cite{Bolton}. This transition is characterised by the onset of elastic response, deriving from the deformation of particles or droplets within a system--spanning contact network that inevitably arises when excluded volume constraints cannot be satisfied. Numerical studies of model athermal systems have revealed similarities to continuous phase transitions: Static quantities such as elastic moduli vanish algebraically as the jamming volume fraction $\phi_{J}$ is approached from above~\cite{Static}, accompanied by diverging length scales in the linear response~\cite{LinRep}. Furthermore, non--linear rheology under continuous shear has revealed a diverging viscosity as $\phi\rightarrow\phi_{J}^{-}$, preconfiguring a finite yield stress above $\phi_{J}$~\cite{Olsson:2007p211,HatanoScaling,Lois}. The flow curves can be made to collapse onto two master curves, one for above the transition and one below, when the stress and strain rates are scaled by powers of $|\phi-\phi_{J}|$~\cite{Olsson:2007p211,HatanoScaling}, analogous to critical scaling functions~\cite{Cardy}.

However, these static or driven systems mask the relaxation mechanisms by which packings reduce unbalanced forces, preventing comparison with the rich phenomenology of relaxation in non--equilibrium systems. The noted similarities with criticality suggest we look there first: Systems quenched to their critical temperature attain local equilibrium on wavelengths shorter than a correlation length $\xi(t)$ that grows algebraically with time~\cite{CriticalAging}. This growing length is at the root of the {\em aging} of correlation and response functions, which include factors of the form $ g[\xi(t_{\rm w}+t)/\xi(t_{\rm w})]$ for a time $t_{\rm w}$ since quench and a lag time~$t$. After suitable normalization the functions $g(x)$ are universal, belonging to a small number of dynamic universality classes. Aging is also present in (non--critical) glassy systems, but $\xi(t)$ is either implicit or has a different physical interpretation such as domain size~\cite{BCKM_Aging}. It is not known if any of this phenomenology carries over to athermal systems at or near $\phi_{J}$.

Here we numerically investigate non--linear relaxation in freely--cooling soft sphere systems for a range of $\phi$ spanning~$\phi_{J}$. We find evidence for a growing length scale $\xi(t)$ corresponding to the relaxation towards {\em mechanical} equilibrium, {\em i.e.} force balance. In analogy with critical scaling, $\xi(t)$ for different $\phi$ can be collapsed onto $\phi<\phi_{J}$ and $\phi>\phi_{J}$ master curves by scaling $\xi$ and $t$ by suitable powers of $|\phi-\phi_{J}|$. From this scaling we also infer an unjamming time that diverges as $\phi\rightarrow\phi_{J}^{-}$. For $\phi>\phi_{J}$, $\xi(t)\sim t$ for all $t$ attributable to elastic propagation, and the particle transport properties obey {\em aging} similar to experiments on a range of soft--matter systems~\cite{Knaebel,Cipelletti,ScatExps}, suggesting a common origin. Finally, for late times we show how $\xi(t)$ maps onto known behavior away from $\phi_{J}$.

%
%

{\em Model.}---We consider viscous soft discs in a square simulation cell of dimensions $L\times L$ with periodic boundary conditions. To reduce ordering effects, the particle diameters $d$ are uniformly distributed over the range $[0.7\langle d\rangle,1.3\langle d\rangle]$ with $\langle d\rangle$ the mean. Particles have equal mass density. Overlapping particles $\alpha$ and $\beta$ with centers separated by a distance $R^{\alpha\beta}<\frac{1}{2}(d^{\alpha}+d^{\beta})$ interact in two ways: {\em (i)}~A~linear repulsive force $f^{\rm el}=\mu[1-2R^{\alpha\beta}/(d^{\alpha}+d^{\beta})]$ acting along the line of centers, and {\em (ii)}~a viscous damping ${\bf f}^{\rm vis}=\eta({\bf v}^{\alpha}-{\bf v}^{\beta})$ which acts to reduce the relative velocity between $\alpha$ and $\beta$. Particles making no contacts move ballistically. Note that for simplicity there is no distinction between tangential and normal velocities in the damping term.

A structureless initial configuration is constructed by depositing particles uniformly over the simulation cell to give the required area fraction $\phi=L^{-2}\sum_{\alpha}\pi(d^{\alpha}/2)^{2}$. Each velocity is initially zero, so the initial energy reservoir is provided by overlapping particles and varies as~$\phi^{2}$. This variation with $\phi$ is weak compared to the much more rapidly varying quantities discussed below, so no crucial dependence on the choice of initial conditions is expected. Particles were iterated using the velocity Verlet algorithm which includes particle inertia~\cite{AllenTildesley}. The simulations were continued until the pressure either changed by less than one part in $10^{5}$ over a time interval corresponding to 20\% of the total, or dropped below some predefined value. For each~$\phi$, the system size $L$ was systematically increased until the pressure and potential energy agreed over the entire time range, to within error bars. We observed that larger system sizes were required as $\phi_{J}$ was approached. Below we normalize lengths by $\langle d\rangle$, times by $t_{0}=\sqrt{\langle d\rangle\langle m\rangle/\mu}$ and the damping coefficient $\eta$ by $\eta_{0}=\sqrt{\mu\langle m\rangle/\langle d\rangle}$, with $\langle m\rangle$ the mean particle mass.

\begin{figure}[htpb]
\includegraphics[width=8.5cm]{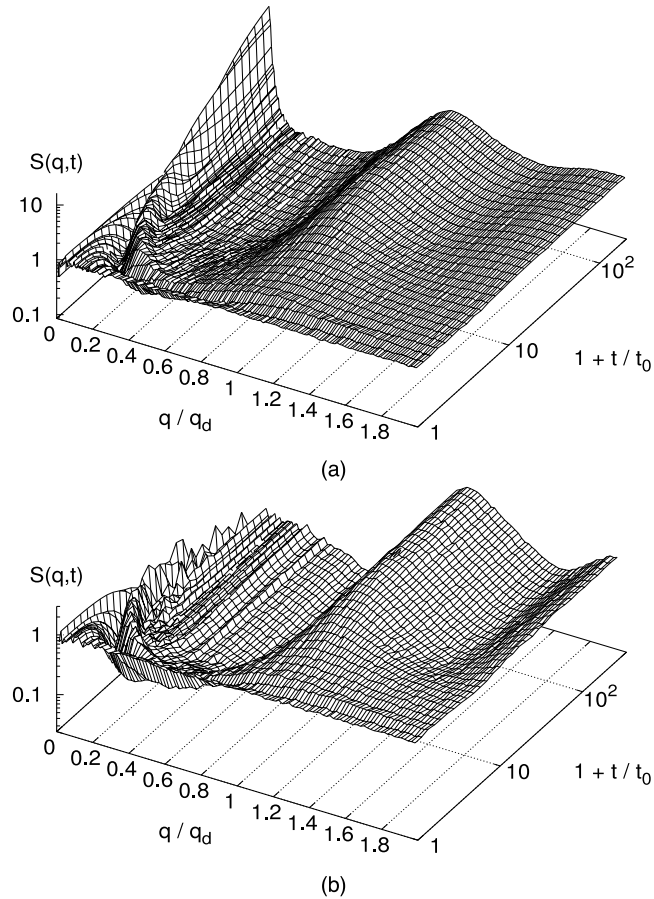}
\caption{The static structure factor $S(q,t)$ versus time for (a)~$\phi=0.7$ and (b)~$\phi=1.0$. The time axis has been scaled by~$t_{0}$ and the $q$--axes  by $q_{\rm d}=2\pi/\langle d\rangle$.}
\label{f:Sq}
\end{figure}

%
%
{\em Results.}---All results below are for a damping coefficient $\eta/\eta_{0}=0.04$, which corresponds to a coefficient of restitution $\approx0.88$ in the dilute limit when there are no multiple contacts. An overview of system evolution is provided in Fig.~\ref{f:Sq}, which shows the static structure factor $S(q,t)=N^{-1}\langle\rho({\bf q},t)\rho({\bf -q},t)\rangle$ for high and low~$\phi$, with $N$ the number of particles and $\rho({\bf q},t)$ the spatially Fourier--transformed number density (here and throughout $\langle\cdots\rangle$ denotes averaging over different initial configurations). Significant $\phi$--dependence is seen only at late times, with the high density system displaying a weak signature of static large--length structure, in contrast to low densities where a peak emerges and grows in height and moves to smaller $q$ with time. This corresponds to the cluster coarsening regime to be discussed later.

\begin{figure}[htpb]
\centering
\includegraphics[width=8.5cm]{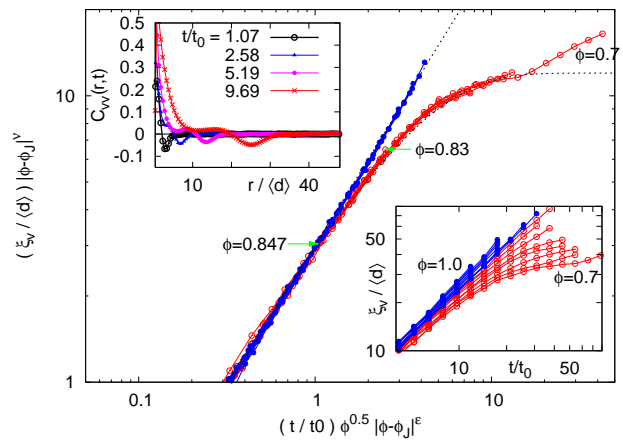}
\caption{{\em (Color online)} Collapse of the velocity correlation lengths $\xi_{\rm v}(t)$ after scaling $\xi_{\rm v}$ by $|\phi-\phi_{J}|^{-\nu}$ and $t$ by$|\phi-\phi_{J}|^{-\rm \epsilon}$, with $\nu=0.57$, $\epsilon=0.6$ and $\phi_{J}=0.843$. The upper line (blue, solid symbols) corresponds to $\phi>\phi_{J}$, the lower line (red, open symbols) to $\phi<\phi_{J}$. The rightmost points of data sets close to $\phi_{J}$ have been indicated. Dashed lines correspond to $\xi_{\rm v}\propto t$ and $\xi_{\rm v}\propto1-{\rm e}^{-t/t_{1}}$ and are intended to guide the eye. {\em (Inset, lower right)} Precollapsed data, left to right for decreasing~$\phi$. Each data set is truncated when $\xi_{\rm v}(t)\approx L/2$. {\em (Inset, upper left)}~Example of $C_{vv}$ for $\phi=0.92$.}
\label{f:lengths}
\end{figure}

For all $\phi$ a local minimum--local maximum pairing emerges at short times and moves to lower $q$ as the system evolves, approximately as $\sim t^{-1}$ for small~$t$. This structural signature of a linearly growing length is also evident in a dynamic length extracted from the same--time velocity correlations $C_{\rm vv}(r=|{\bf x}^{\alpha}-{\bf x}^{\beta}|,t)={\mathcal N} \langle\sum_{\alpha,\beta}{\bf v}({\bf x}^{\alpha},t)\cdot{\bf v}({\bf x}^{\beta},t)\rangle$, normalized so that $C_{\rm vv}(0,t)\equiv1$. Following Olsson and Teitel~\cite{Olsson:2007p211} we identify the characteristic velocity correlation length $\xi_{\rm v}(t)$ with the global minimum of $C_{\rm vv}$\,. The growth of $\xi_{\rm v}(t)$ for different $\phi$, and examples of $C_{vv}$, are given in insets to Fig.~\ref{f:lengths}.

The data for all $\phi$ can be collapsed onto two master curves by scaling $\xi_{\rm v}(t)$ by $|\phi-\phi_{J}|^{-\nu}$ and $t$ by $|\phi-\phi_{J}|^{-\epsilon}$, with $\nu=0.57\pm0.05$, $\epsilon=0.6\pm0.05$ and $\phi_{J}=0.843\pm0.001$, as demonstrated in Fig.~\ref{f:lengths}. Note that we also scale $t$ by a non--critical factor $\phi^{1/2}$ to improve collapse at small times, in the spirit of corrections to scaling~\cite{Cardy}, but this does not alter the exponents. The exponent $\nu$ is consistent with that already found for steady flow~\cite{Olsson:2007p211}, but as for that protocol, we cannot achieve reasonable collapse using the exponent $\approx0.5$ for the diverging length in linear response~\cite{LinRep}, and conclude these two lengths are unrelated. Indeed, for all $\phi>\phi_{J}$, $\xi_{\rm v}(t)\propto t$ and hence diverges with time, whereas the linear response lengths only diverge for $\phi\rightarrow\phi_{J}^{+}$. (Convergence of {\em e.g.} pressure with system size is achieved as long as the dynamics and statics have decoupled by the time $\xi_{\rm v}\approx L/2$).

\begin{figure}[htpb]
\centering
\includegraphics[width=8.5cm]{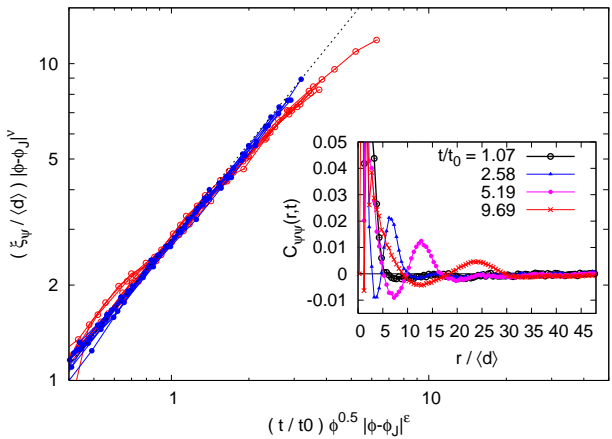}
\caption{{\em (Color online)} Characteristic length for force balance $\xi_{\psi}(t)$ versus time under the same scaling as Fig.~\ref{f:lengths} with the same notation. The dashed line is $\xi_{\psi}(t)\propto t$. {\em (Inset)}~$C_{\psi\psi}(r,t)={\mathcal N}(\overline{\psi(0)\psi(r)}-\bar{\psi}^{2})$
at different times for $\phi=0.92$. $\xi_{\psi}(t)$ is identified with the rightmost maximum.}
\label{f:force_bal}
\end{figure}

Instead we propose that $\xi_{\rm v}(t)$ corresponds to the length over which the system approaches mechanical equilibrium, {\em i.e.} force balance. To test this hypothesis, we assign to each particle $\alpha$ the scalar quantity $\psi^{\alpha}=1-|\sum_{\beta}{\bf f}^{\alpha\beta}|/\sum_{\beta}|{\bf f}^{\alpha\beta}|$, where the sums are over all particles $\beta$ in contact with $\alpha$ and  ${\bf f}^{\alpha\beta}$ is the {\em elastic--only} component of the corresponding interaction force. Note that $\sum_{\beta}{\bf f}^{\alpha\beta}$ is the resultant elastic force and $\sum_{\beta}|{\bf f}^{\alpha\beta}|$ a normalization factor, so higher $\psi$ means more balanced forces, with perfect balance at $\psi=1$. Proceeding as before, we measure spatial correlations $C_{\psi\psi}(r,t)$ in $\psi({\bf x}^{\alpha})=\psi^{\alpha}$ and extract a characteristic length $\xi_{\psi}(t)$, as described in Fig.~\ref{f:force_bal}. As with the velocity correlation length, $\xi_{\psi}$ initially grows linearly and obeys the same scaling with $|\phi-\phi_{J}|$ as for $\xi_{\rm v}$, confirming they both reflect the same relaxation process, although $\xi_{\psi}$ is statistically noisy and the reliable time window correspondingly smaller. Nonetheless this confirms that mechanical equilibrium is reached on a growing length $\xi(t)\sim\xi_{\rm v}(t)\sim\xi_{\psi}(t)$.

The master curves themselves give insights into the physics underlying relaxation. For short times $\xi_{\rm v}(t)\propto t$ for all $\phi$, which, since there is no ballistic transport of particles {\em (see below)}, must be due to {\em elastic} propagation through multi--particle contact networks. Inverting the axes scaling gives a speed of sound $c\propto|\phi-\phi_{J}|^{\epsilon-\nu}$, which is finite at $\phi_{J}$ when $\epsilon\equiv\nu$, which is within error bars. Below~$\phi_{J}$, $\xi_{\rm v}(t)$ also initially grows linearly before saturating at a finite value, at which point the data collapse fails and cluster coarsening (which is {\em not} controlled by $\phi_{J}$) begins. The allows us to define an `unjamming' time when the plateau is reached, which by reversing the axes scaling is seen to diverge as $\sim(\phi_{J}-\phi)^{-\epsilon}$.

\begin{figure}[htpb]
\centering
\includegraphics[width=8.5cm]{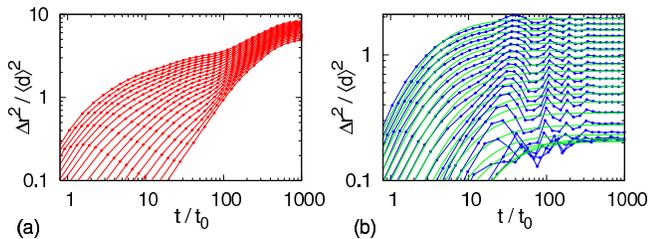}
\caption{{\em (Color online)} Mean squared displacement $\Delta r^{2}(t_{w}+t,t_{\rm w})$ for single runs at (a)~$\phi=0.8<\phi_{J}$ and (b)~$\phi=0.88>\phi_{J}$. In both cases, lines from top to bottom correspond to geometrically increasing $t_{\rm w}$ in the range $0.4<t_{w}/t_{0}<40$. In (b) the smooth green lines are smoothing fits (see text).}
\label{f:msd_plateau}
\end{figure}

%
%
{\em Aging.---} If $\xi(t)$ controls the relaxation as claimed, the mean--squared displacement (MSD) $\Delta r^{2}(t_{\rm w}+t,t_{\rm w})=N^{-1}\sum_{\alpha}|{\bf x}^{\alpha}(t_{\rm w}+t)-{\bf x}^{\alpha}(t_{\rm w})|^{2}$ should be a function of $\xi(t_{\rm w}+t)/\xi(t_{\rm w})$~\cite{BCKM_Aging}. For $\phi>\phi_{J}$, $\xi(t)\propto t$ so we expect $\Delta r^{2}(t_{\rm w}+t,t_{\rm w})\sim g(t/t_{\rm w})$ corresponding to full aging. Conversely, $\xi(t)$ approaches a $\phi$--dependent constant for $\phi<\phi_{J}$ and critical aging should cease (although some other form of aging may recur deep in the cluster coarsening regime). Examples of $\Delta r^{2}(t_{\rm w}+t,t_{\rm w})$ above and below $\phi_{J}$ are given in Fig.~\ref{f:msd_plateau}. For $\phi>\phi_{J}$ the MSD takes a fixed form which systematically scales to lower amplitudes and later times with increasing $t_{\rm w}$. By contrast, for $\phi<\phi_{J}$ no such systematic scaling is apparent over the available time window.

\begin{figure}[htpb]
\centering
\includegraphics[width=8.5cm]{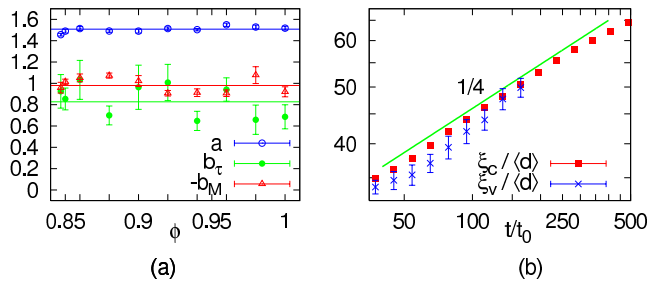}
\caption{{\em (Color online)} 
{\em (a)}~Transport and aging exponents versus density. Smooth lines are the average over all~$\phi$.
{\em (b)}~Characteristic cluster size $\xi_{c}$ and velocity correlation length $\xi_{\rm v}$ at late times for $\phi=0.7$.
}
\label{f:exponents}
\end{figure}

Aging has been experimentally observed in a range of relaxing soft--matter systems~\cite{Knaebel,Cipelletti,ScatExps}, and we speculate that the underlying mechanism may be the same as in this athermal system. To test this we must first quantify the observed $t_{\rm w}$--scaling for $\phi>\phi_{J}$. We first smooth the data by fitting each $t_{\rm w}$--curve to the 3--parameter fit $M(t_{\rm w})/\{1+[t/\tau(t_{\rm w})]^{-\diffexp}\}$, as shown in Fig.~\ref{f:msd_plateau}(b). $M(t_{\rm w})$ describes the variation in the overall amplitude of particle transport with~$t_{\rm w}$, $\tau(t_{\rm w})$ is a relaxation time, and $a$ is the early--time growth exponent. For large $t_{\rm w}$ both $M(t_{\rm w})$ and $\tau(t_{\rm w})$ are expected to scale algebraically with $t_{\rm w}$, so we fit each to the form $A(1+t_{\rm w}/B)^{\msdexp,\tauexp}$ resp., to extract $\msdexp$ and $\tauexp$. The exponents $a$, $\msdexp$ and $\tauexp$ for each $\phi$ are plotted in Fig.~\ref{f:exponents}(a). Since there is little apparent variation with $\phi$, we can improve the statistics by averaging over all $\phi$, giving $a=1.51\pm0.01$, $\msdexp=-0.98\pm0.02$ and $\tauexp=0.84\pm0.05$. This latter value would appear to suggest {\em subaging} rather than the full aging  $\tauexp\equiv1$ expected for a critical point~\cite{CriticalAging}, but we cannot yet rule out systematic errors due to the chosen fitting functions.

It is now possible to compare these findings to the equivalent quantities measured in the experiments~\cite{Knaebel,Cipelletti,ScatExps}. We find three areas of agreement: {\em (i)}~Superdiffusive particle transport $\Delta r^{2}\sim t^{a}$ with $a\approx1.5$, as inferred from the speckle decay in experiments~\cite{Knaebel} and directly measured here [for $t\ll\tau(t_{\rm w})$]; {\em (ii)}~Aging, with experimental $\tauexp$ in the range 0.77 to~1.8, compared to $\tauexp\approx0.84$ here, and {\em (iii)}~Convective decay of the scattering vector, which has been interpreted as evidence for the ballistic motion of elastic strain deformations through the material~\cite{Cipelletti,Bouchaud:2002p19}. Elastic waves are also present in our system, and while we do not refute this interpretation of the data, it is interesting to note that we observe a linearly--growing length in this model, namely the correlation length $\xi(t)\sim t$ detailed above. We hypothesise that this may be the true origin of the experimentally--observed ballistic growth law. In this context, closer comparison with experimental data would be desirable.

%
%
{\em Late times.---} For $\phi>\phi_{J}$, $\xi(t)$ becomes arbitrarily large and the system reaches global mechanical equilibrium. The scaling properties of these static systems have been studied by non--inertial algorithms~\cite{Static,LinRep}. For comparison, in Table~\ref{t:scaling} we give the corresponding quantities for states generated by our inertial algorithm. The agreement with previous results confirms the robustness of the static state to the preparation procedure. As our density window is somewhat broad, extending to roughly $15\%$ either side of~$\phi_{J}$, simple scaling cannot be assumed~\cite{Cardy} and a correction to scaling was included when fitting.



 \begin{table}
\caption{\label{t:scaling}Fits of dimensionless pressure, P.E., excess coordination number, relative particle overlap and shear modulus at $t=\infty$ to $A(\phi-\phi_{J})^{\alpha}[1+B(\phi-\phi_{J})]$, with a correction to scaling $\propto B$. Numbers in brackets denote uncertainty in the last digit. The bulk elastic modulus remains finite as $\phi\rightarrow\phi_{J}^{+}$. }
\begin{ruledtabular}
\begin{tabular}{c@{\quad}|@{\quad}ccc}
Quantity & Prefactor $A$ & $\phi_{J}$ & Exponent $\alpha$
\\
\hline
Pressure & 0.9(1) & 0.8433(4) & 1.03(4) \\
Potential Energy & 0.05(1) & 0.8434(4) & 1.98(9) \\
$z-z^{\rm iso}$ & 4.3(4) & 0.8432(4) & 0.55(4) \\
Particle overlap & 0.4(1) & 0.8433(5) & 1.03(4) \\
$G/G_{\rm affine}$ & 4.8(4) & 0.844(1) & 0.47(3) \\
\end{tabular}
\end{ruledtabular}
\end{table}


%
%

For $\phi<\phi_{J}$, the system unjams and an unstable coupling between dissipation and density fluctuations leads to system--wide mass separation into clusters and voids. In this regime the small--$q$ peak in $S(q,t)$ is pronounced, allowing a characteristic cluster length $\xi_{\rm c}(t)$ to be extracted, as plotted in Fig.~\ref{f:exponents}(b). The velocity correlation length $\xi_{\rm v}$ is also shown, and is similar to $\xi_{\rm c}$ where their time windows overlap, demonstrating that $\xi(t)$ tracks cluster growth after unjamming. Over the available data window $\xi_{\rm c}(t)\approx t^{0.26\pm0.03}$, towards the lower end of quoted values for hard spheres~\cite{GranCool}, suggesting that subsequent evolution is described by granular gas theory.

{\em Conclusions.---} The relaxation of freely cooling athermal systems considered here support the view that the model `point--J' jamming transition is closer in nature to a continuous phase transition than a glass transition: While aging is common to both, the critical scaling of $\xi(t)$~in Fig.~\ref{f:lengths}, controlled by a single point~$\phi_{J}$, is expected only for critical points. The correspondence is of course not complete, and as for continuous shear~\cite{HatanoScaling} we expect the master curves to depend on the interaction potential. Nonetheless we believe a fundamental understanding of this important transition will best approached from the standpoint of critical point theory, and further modelling in this direction would be desirable.

{\em Acknowledgements.---} The author would like to acknowledge H. Tanaka for useful discussions.


\begin{thebibliography}{31}
\expandafter\ifx\csname natexlab\endcsname\relax\def\natexlab#1{#1}\fi
\expandafter\ifx\csname bibnamefont\endcsname\relax
  \def\bibnamefont#1{#1}\fi
\expandafter\ifx\csname bibfnamefont\endcsname\relax
  \def\bibfnamefont#1{#1}\fi
\expandafter\ifx\csname citenamefont\endcsname\relax
  \def\citenamefont#1{#1}\fi
\expandafter\ifx\csname url\endcsname\relax
  \def\url#1{\texttt{#1}}\fi
\expandafter\ifx\csname urlprefix\endcsname\relax\def\urlprefix{URL }\fi
\providecommand{\bibinfo}[2]{#2}
\providecommand{\eprint}[2][]{\url{#2}}

\bibitem{Bolton}
\bibinfo{author}{\bibfnamefont{F.}~\bibnamefont{Bolton}} \bibnamefont{and}
  \bibinfo{author}{\bibfnamefont{D.}~\bibnamefont{Weaire}},
  \bibinfo{journal}{Phys. Rev. Lett.} \textbf{\bibinfo{volume}{65}},
  \bibinfo{pages}{3449} (\bibinfo{year}{1990});
\bibinfo{author}{\bibfnamefont{E.}~\bibnamefont{Aharonov}} \bibnamefont{and}
  \bibinfo{author}{\bibfnamefont{D.}~\bibnamefont{Sparks}},
  \bibinfo{journal}{Phys. Rev. E} \textbf{\bibinfo{volume}{60}},
  \bibinfo{pages}{6890} (\bibinfo{year}{1999});
\bibinfo{author}{\bibfnamefont{T.}~\bibnamefont{Mason}},
  \bibinfo{author}{\bibfnamefont{J.}~\bibnamefont{Bibette}}, \bibnamefont{and}
  \bibinfo{author}{\bibfnamefont{D.}~\bibnamefont{Weitz}},
  \bibinfo{journal}{Phys. Rev. Lett.} \textbf{\bibinfo{volume}{75}},
  \bibinfo{pages}{2051} (\bibinfo{year}{1995}).

\bibitem{Static}
\bibinfo{author}{\bibfnamefont{C.~S.} \bibnamefont{O'Hern}},
  \bibinfo{author}{\bibfnamefont{L.~E.} \bibnamefont{Silbert}},
  \bibnamefont{and} \bibinfo{author}{\bibfnamefont{S.~R.} \bibnamefont{Nagel}},
  \bibinfo{journal}{Phys. Rev. E} \textbf{\bibinfo{volume}{68}},
  \bibinfo{pages}{011306} (\bibinfo{year}{2003});
\bibinfo{author}{\bibfnamefont{C.~S.} \bibnamefont{O'Hern}},
  \bibinfo{author}{\bibfnamefont{S.}~\bibnamefont{Langer}},
  \bibinfo{author}{\bibfnamefont{A.}~\bibnamefont{Liu}}, \bibnamefont{and}
  \bibinfo{author}{\bibfnamefont{S.~R.} \bibnamefont{Nagel}},
  \bibinfo{journal}{Phys. Rev. Lett.} \textbf{\bibinfo{volume}{88}},
  \bibinfo{pages}{075507} (\bibinfo{year}{2002}).

\bibitem{LinRep}
\bibinfo{author}{\bibfnamefont{L.~E.} \bibnamefont{Silbert}} {\em et al.},
  \bibinfo{journal}{Phys. Rev. Lett.} \textbf{\bibinfo{volume}{95}},
  \bibinfo{pages}{098301} (\bibinfo{year}{2005});
\bibinfo{author}{\bibfnamefont{M.}~\bibnamefont{Wyart}},
  \bibinfo{journal}{Phys. Rev. E} \textbf{\bibinfo{volume}{72}},
  \bibinfo{pages}{051306} (\bibinfo{year}{2005});
\bibinfo{author}{\bibfnamefont{M.}~\bibnamefont{Wyart}},
  \bibinfo{author}{\bibfnamefont{S.~R.} \bibnamefont{Nagel}}, \bibnamefont{and}
  \bibinfo{author}{\bibfnamefont{T.~A.} \bibnamefont{Witten}},
  \bibinfo{journal}{Europhys Lett} \textbf{\bibinfo{volume}{72}},
  \bibinfo{pages}{486} (\bibinfo{year}{2007});
\bibinfo{author}{\bibfnamefont{W.~G.} \bibnamefont{Ellenbroek}},
  \bibinfo{author}{\bibfnamefont{E.}~\bibnamefont{Somfai}},
  \bibinfo{author}{\bibfnamefont{M.~V.} \bibnamefont{Hecke}}, \bibnamefont{and}
  \bibinfo{author}{\bibfnamefont{W.~V.} \bibnamefont{Saarloos}},
  \bibinfo{journal}{Phys. Rev. Lett.} \textbf{\bibinfo{volume}{97}},
  \bibinfo{pages}{258001} (\bibinfo{year}{2006});
\bibinfo{author}{\bibfnamefont{J.~A.} \bibnamefont{Drocco}} {\em et al.},
  \bibinfo{journal}{Phys. Rev. Lett.} \textbf{\bibinfo{volume}{95}},
  \bibinfo{pages}{088001} (\bibinfo{year}{2005}).

\bibitem[{\citenamefont{Olsson and Teitel}(2007)}]{Olsson:2007p211}
\bibinfo{author}{\bibfnamefont{P.}~\bibnamefont{Olsson}} \bibnamefont{and}
  \bibinfo{author}{\bibfnamefont{S.}~\bibnamefont{Teitel}},
  \bibinfo{journal}{Phys. Rev. Lett.} \textbf{\bibinfo{volume}{99}},
  \bibinfo{pages}{178001} (\bibinfo{year}{2007}).

\bibitem[{\citenamefont{Hatano}(2008)}]{HatanoScaling}
\bibinfo{author}{\bibfnamefont{T.}~\bibnamefont{Hatano}}
  (\bibinfo{year}{2008}), \bibinfo{note}{cond-mat/0803.2296v3}.

\bibitem{Lois}
\bibinfo{author}{\bibfnamefont{G.}~\bibnamefont{Lois}},
  \bibinfo{author}{\bibfnamefont{A.}~\bibnamefont{Lemaitre}}, \bibnamefont{and}
  \bibinfo{author}{\bibfnamefont{J.~M.} \bibnamefont{Carlson}},
  \bibinfo{journal}{Europhys Lett} \textbf{\bibinfo{volume}{76}},
  \bibinfo{pages}{318} (\bibinfo{year}{2006});
\bibinfo{author}{\bibfnamefont{G.}~\bibnamefont{Lois}} \bibnamefont{and}
  \bibinfo{author}{\bibfnamefont{J.~M.} \bibnamefont{Carlson}},
  \bibinfo{journal}{Europhys Lett} \textbf{\bibinfo{volume}{80}},
  \bibinfo{pages}{58001} (\bibinfo{year}{2007}).

\bibitem[{\citenamefont{Cardy}(1996)}]{Cardy}
\bibinfo{author}{\bibfnamefont{J.}~\bibnamefont{Cardy}},
  \emph{\bibinfo{title}{Scaling and Renomalization in Statistical Physics}}
  (\bibinfo{publisher}{Cambridge University Press},
  \bibinfo{address}{Cambridge}, \bibinfo{year}{1996}).

\bibitem{CriticalAging}
\bibinfo{author}{\bibfnamefont{L.}~\bibnamefont{Berthier}},
  \bibinfo{author}{\bibfnamefont{P.}~\bibnamefont{Holdsworth}},
  \bibnamefont{and} \bibinfo{author}{\bibfnamefont{M.}~\bibnamefont{Sellitto}},
  \bibinfo{journal}{J Phys A-Math Gen} \textbf{\bibinfo{volume}{34}},
  \bibinfo{pages}{1805} (\bibinfo{year}{2001});
\bibinfo{author}{\bibfnamefont{S.}~\bibnamefont{Abriet}} \bibnamefont{and}
  \bibinfo{author}{\bibfnamefont{D.}~\bibnamefont{Karevski}},
  \bibinfo{journal}{Eur Phys J B} \textbf{\bibinfo{volume}{37}},
  \bibinfo{pages}{47} (\bibinfo{year}{2004});
\bibinfo{author}{\bibfnamefont{P.}~\bibnamefont{Calabrese}} \bibnamefont{and}
  \bibinfo{author}{\bibfnamefont{A.}~\bibnamefont{Gambassi}},
  J. Phys. A {\bf 38}, R133 (2005);
\bibinfo{author}{\bibfnamefont{P.}~\bibnamefont{Calabrese}},
  \bibinfo{author}{\bibfnamefont{A.}~\bibnamefont{Gambassi}}, \bibnamefont{and}
  \bibinfo{author}{\bibfnamefont{F.}~\bibnamefont{Krzakala}},
  \bibinfo{journal}{J Stat Mech} \bibinfo{pages}{P06016}
  (\bibinfo{year}{2006}).

\bibitem[{\citenamefont{Bouchaud et~al.}(1997)\citenamefont{Bouchaud,
  Cugliandolo, Kurchan, and M\'ezard}}]{BCKM_Aging}
\bibinfo{author}{\bibfnamefont{J.-P.} \bibnamefont{Bouchaud}},
  \bibinfo{author}{\bibfnamefont{L.~F.} \bibnamefont{Cugliandolo}},
  \bibinfo{author}{\bibfnamefont{J.}~\bibnamefont{Kurchan}}, \bibnamefont{and}
  \bibinfo{author}{\bibfnamefont{M.}~\bibnamefont{M\'ezard}}, in
  \emph{\bibinfo{booktitle}{Slow relaxations and non--equilibrium dynamics in
  condensed matter}}, edited by \bibinfo{editor}{\bibfnamefont{A.~P.}
  \bibnamefont{Young}} (\bibinfo{publisher}{World Scientific},
  \bibinfo{address}{Singapore}, \bibinfo{year}{1997}).

\bibitem{Knaebel}
\bibinfo{author}{\bibfnamefont{A.}~\bibnamefont{Knaebel}}
{\em et al.},
  \bibinfo{journal}{Europhys Lett} \textbf{\bibinfo{volume}{52}},
  \bibinfo{pages}{73} (\bibinfo{year}{2000}).

\bibitem{Cipelletti}
\bibinfo{author}{\bibfnamefont{L.}~\bibnamefont{Cipelletti}}
{\em et al.},
  \bibinfo{journal}{Phys. Rev. Lett.} \textbf{\bibinfo{volume}{84}},
  \bibinfo{pages}{2275} (\bibinfo{year}{2000}).

\bibitem{ScatExps}
\bibinfo{author}{\bibfnamefont{R.}~\bibnamefont{Bandyopadhyay}}
{\em et al.},
  \bibinfo{journal}{Phys. Rev. Lett.} \textbf{\bibinfo{volume}{93}},
  \bibinfo{pages}{228302} (\bibinfo{year}{2004});
\bibinfo{author}{\bibfnamefont{M.}~\bibnamefont{Bellour}}
{\em et al.},
  \bibinfo{journal}{Phys. Rev. E} \textbf{\bibinfo{volume}{67}},
  \bibinfo{pages}{031405} (\bibinfo{year}{2003}).



\bibitem[{\citenamefont{Allen and Tildesley}(1987)}]{AllenTildesley}
\bibinfo{author}{\bibfnamefont{M.~P.} \bibnamefont{Allen}} \bibnamefont{and}
  \bibinfo{author}{\bibfnamefont{D.~J.} \bibnamefont{Tildesley}},
  \emph{\bibinfo{title}{Computer Simulation of Liquids}}
  (\bibinfo{publisher}{Oxford University Press}, \bibinfo{address}{Oxford},
  \bibinfo{year}{1987}).


\bibitem[{\citenamefont{Bouchaud and Pitard}(2002)}]{Bouchaud:2002p19}
\bibinfo{author}{\bibfnamefont{J.}~\bibnamefont{Bouchaud}} \bibnamefont{and}
  \bibinfo{author}{\bibfnamefont{E.}~\bibnamefont{Pitard}},
  \bibinfo{journal}{Eur Phys J E} \textbf{\bibinfo{volume}{9}},
  \bibinfo{pages}{287} (\bibinfo{year}{2002}).

\bibitem{GranCool}
\bibinfo{author}{\bibfnamefont{S.~R.} \bibnamefont{Ahmad}} \bibnamefont{and}
  \bibinfo{author}{\bibfnamefont{S.}~\bibnamefont{Puri}},
  \bibinfo{journal}{Phys. Rev. E} \textbf{\bibinfo{volume}{75}},
  \bibinfo{pages}{031302} (\bibinfo{year}{2007});
\bibinfo{author}{\bibfnamefont{C.}~\bibnamefont{Cattuto}} \bibnamefont{and}
  \bibinfo{author}{\bibfnamefont{U.}~\bibnamefont{Marconi}},
  \bibinfo{journal}{Phys. Rev. Lett.} \textbf{\bibinfo{volume}{92}},
  \bibinfo{pages}{174502} (\bibinfo{year}{2004});
\bibinfo{author}{\bibfnamefont{S.}~\bibnamefont{Luding}} \bibnamefont{and}
  \bibinfo{author}{\bibfnamefont{H.}~\bibnamefont{Herrmann}},
  \bibinfo{journal}{Chaos} \textbf{\bibinfo{volume}{9}}, \bibinfo{pages}{673}
  (\bibinfo{year}{1999});
\bibinfo{author}{\bibfnamefont{S.~K.} \bibnamefont{Das}} \bibnamefont{and}
  \bibinfo{author}{\bibfnamefont{S.}~\bibnamefont{Puri}},
 \bibinfo{journal}{Phys. Rev. E} \textbf{\bibinfo{volume}{68}},
 \bibinfo{pages}{011302} (\bibinfo{year}{2003}).


\end{thebibliography}


\end{document}